\documentclass{JHEP}
\usepackage{epsfig}

\title{Brane World Sum Rules}

\author{Gary Gibbons\\
    DAMTP, CMS, Cambridge Univ., Wilberforce Road, Cambridge CB3 0WA,
UK\\ E-mail: \email{G.W.Gibbons@damtp.cam.ac.uk}}
\author{Renata Kallosh\\
Department of Physics, Stanford University, Stanford, CA 94305, USA\\
    E-mail: \email{kallosh@stanford.edu}}
\author{Andrei Linde\\
    Department of Physics, Stanford University, Stanford, CA 94305,
USA\\
    E-mail: \email{linde@physics.stanford.edu}}
\received{\today}       %%
 \preprint{SU-ITP-00-30\\ DAMTP-2000-130\\ \hepth{0011225}\\ \today}

\abstract{A set of consistency conditions is derived from Einstein equations
for brane world scenarios with a spatially periodic internal space. In particular, the sum of the total tension of the flat branes and the non-negative integral of the gradient energy of the bulk scalars must vanish. This constraint allows us to make a simple consistency check of several models. We show that the two-brane Randall-Sundrum model satisfies this constraint, but it does not allow a generalization with smooth branes (domain walls), independently of the issue of supersymmetry. The Goldberger-Wise
model of brane stabilization has to include the backreaction on the metric and the fine tuning of the cosmological constant to satisfy the constraints. We check that this is achieved in the DeWolfe-Freedman-Gubser-Karch scenario. Our constraints are automatically satisfied in supersymmetric brane world models.}


\begin{document}

\section{Introduction}
 There is a strong  interest in the possibility that our universe can be
represented as a brane in a higher dimensional space \cite{savas,RSI}.
The number of new brane world models is growing fast. Quite often it is
difficult to make any judgement concerning the  consistency of these
models, not only because they are rather complicated, but also because of
many `plausible' assumptions and approximations made by their authors.
One of the greatest problems is that one must ensure that the effective
cosmological constant on the brane is smaller than $10^{-120}$ in the
Planck units.

Therefore it would be quite useful to have a set of simple and
sufficiently general rules which would allow one to test  new models.
In this paper we will derive a set of consistency conditions which can
be helpful in this respect.  We  will consider  some combinations
of the  Einstein equations for  warped metrics in the theory of
gravity interacting with a collection of scalar fields. The simple case
is the one with an internal  manifold without boundary, on which the
integral of the total derivative vanish, as when the internal space is
spatially periodic. This leads to a number of conditions, particularly for
the brane tensions and the gradient energy of scalars.  Some of these
conditions have been derived before, see e.g. \cite{gary,ellwanger,peter},
but the most important constraint that we are going to use is new. It shows
that the sum of the total tension of the branes and the non-negative
integral  of the gradient energy of scalars is proportional to the
four-dimensional curvature of
the branes. For flat branes, corresponding to empty space with zero
cosmological constant, this sum must vanish.

This condition allows us to make a simple consistency check of several
popular versions of the brane world scenario. We show, in particular,
that whereas the original Randall-Sundrum model with two singular branes
\cite{RSI} satisfies this condition, it does not have a consistent
generalization with smooth branes (domain walls), independently of the
issue of supersymmetry. We also show that the Goldberger-Wise brane
stabilization mechanism \cite{GW} has to include the backreaction of the
scalar field on the metric and a fine tuning of the cosmological constant
to satisfy the constraints.\footnote{For the sake of clarity
and to  avoid the possibility of confusion,
we point out that for the purposes of this paper
"fine tuning"  refers to the need to constrain one or more  parameters
appearing in the classical equations of motion, or equivalently 
in the classical action functional,  in order to obtain
a solution of the desired type.  The larger the number of parameters
the larger the degree of fine tuning. If no solution exists for any
physically sensible choice of parameters we  say that the
model is inconsistent. Although one may use perturbation theory to check for fine tuning this is merely because exact solutions are not available.
The idea of fine tuning is exact and does not rest on
perturbation theory. Finally, although the considerations of this
and related papers  may throw some light on the full quantum mechanical  problem, they operate at best with postulated forms of a classical  effective action. The relationship between the parameters appearing in the classical
effective action and the  parameters appearing in some underlying
`fundamental action,'  not least because the models are
non-renormalizable, remains  obscure. The assumption
 that fine-tuning of effective parameters requires
fine tuning of fundamental parameters may or may not be correct.
For example in supersymmetric actions it may follow  automatically.}
 We will explain how fine tuning
is achieved in the
DeWolfe-Freedman-Gubser-Karch model  \cite{DFGK}. We also verify that this
consistency condition is automatically satisfied in the supersymmetric brane
world scenario proposed by Bergshoeff, Kallosh, and Van Proeyen \cite{BKP}.
The flatness of the branes in this cases follows from supersymmetry of the
bulk and brane action.

%%%%%%%%%%%%%%%%%%%%%%%%%%%%%%%%%%%%%%%%%%%%%%%%%%%%%%%%%%%%
\section{Derivation of Sum Rules from Einstein Equations}
 Here we present the   {\it basic equations for  warp factors and
compactifications}. Consider the $D$-dimensional warped product metric
of a $p+1$ dimensional
spacetime with coordinates $x^\mu$ with a $D-p-1$ "internal" space with
coordinates $y^m$
\begin{equation}
ds^2=g_{mn}(y)\, dy^m dy^n + W^2(y)\, g_{\mu \nu}(x)\, dx^\mu dx^\nu ~.
\label{warped}
\end{equation}
Then the Ricci tensor components of the bulk are related to those in
spacetime and the internal space by the formulae
\begin{equation}
   ^{(D)}R_{\mu \nu}=^{(p+1)}R_{\mu \nu}
 -{g_{\mu \nu} \over p+1} { 1 \over W^{p-1}} \nabla ^2 (W^{p+1})
\label{first}
\end{equation}
\begin{equation}
   ^{(D)}R_{mn}=^{(D-p-1)}R_{mn}-{p+1 \over W} \nabla_{m} \nabla_{n} W~,
\label{second}
\end{equation}
where $\nabla_m$ and $\nabla ^2$ are respectively the covariant
derivative and Laplacian with respect
 to the internal metric $g_{mn}$.
The Einstein equations in the bulk are
\begin{equation}
   ^{(D)}R_{MN}= 8\pi G_D \left(
T_{MN}-{ 1 \over D-2} g_{MN} T^C_C \right), \label{einstein}
\end{equation}
 where $T_{MN}$ is the energy
momentum tensor of the matter and $G_D$ is the bulk Newton's constant.

If the internal space is ``closed,''   i.e.  compact without boundary,
then by assuming some general properties of the energy momentum tensor,
multiplying the equations by suitable functions of the warp factor $W$
and integrating over the internal space one obtains some necessary
conditions for compactification with a flat (or curved) spacetime.

 In many cases the singular sources are introduced at $y=y_i$ by
adding  brane actions to the bulk action of the form:
\begin{equation}
  S_{br} = -\sum_{\alpha} \int d^{p+1} x \sqrt {-\det
  g_{\mu\nu}}\; \lambda_\alpha (\Phi)~.
\label{brane}
\end{equation}
The curvature as well as the  energy-momentum tensor will have some
singular contributions of the form $\sim \delta (y-y_i)$. We will
consider examples below.

%%%%%%%%%%%%%%%%%%%%%%%%%%%%%%%%%%%%%%%%%%%%%%%%%%%%%%%

A {\it Special case of $D=5$, $p=3$ }
with flat
 internal   metric $g_{yy} =1$ is relevant for domain walls in a
5-dimensional spacetime. The basic equations reduce to
\begin{equation}
   ^{(5)}R_{\mu \nu}=^{(4)}R_{\mu \nu}
 -{\eta_{\mu \nu} \over 4} { 1 \over W^{2}}  (W^{4})^{''}
\label{5first}
\end{equation}
\begin{equation}
   ^{(5)}R_{55}=-{4\over W}  W^{''}~.
\label{5second}
\end{equation}
We work with mostly + metric and the $'$ stands for derivative with
respect to $y$.

 We may rewrite these
equations as follows
\begin{eqnarray}\label{basic1}
 ^{(5)}R_\mu {}^\mu - R_{g}  W^{-2}  & =  &
-12(W')^2 W^{-2}  -4 W^{''}W^{-1}~, \\ \nonumber\\
 ^{(5)} R_5{}^5 &=&-4
W^{''} W^{-1} ~.
\label{basic2}
\end{eqnarray}
Here we have introduced the
notation $R_g$ for the curvature scalar of the 4-dimensional space
with
the metric $g_{\mu\nu}$.
\begin{equation}
   ^{(4)}R_\mu {}^\mu= R_{g}~.
\label{4curv}
\end{equation}
From eqs. (\ref{basic1}), (\ref{basic2}),
the bulk curvature $ ^{(5)}R$ is related to the brane curvature by
\begin{equation}
  ^{(5)}R= R_g W^{-2}-12(W')^2 W^{-2}  -8 W^{''}W^{-1}~.
\label{Rrelation}
\end{equation}
One may  take  linear combination of Eq. (\ref{basic1}) multiplied by
$(1-n)W^n$ and Eq. (\ref{basic2}) multiplied by $(n-4)W^n$  to obtain:
\begin{equation}
\left  ( A' e^{n A}\right) ' = {(W^n)^{''}\over n}= {1-n\over 12} W^n
\left(^{(5)} R_\mu {}^\mu- R_g W^{-2}\right) +{n-4 \over 12} W^n ~
^{(5)} R_5{}^5 ~,
\label{comb}
\end{equation}
where $W= e^A$.
 Using  the Einstein eqs.
$^{(5)}R_{AB}= 8\pi G_5 \bigl (
T_{AB}-{ 1 \over 3} g_{AB} T^C_C \bigr )$,
we  can express the desired
combination of Ricci  tensors in terms of  the energy-momentum tensor.
We note
that
${1\over 8\pi G_5} R_\mu {}^\mu= -{1\over 3}T_\mu {}^\mu - {4\over 3}T_5
{}^5$ and ${1\over 8\pi G_5} R_5{}^5= -{1\over 3}T_\mu {}^\mu +{2\over
3}T_5 {}^5$.
Thus for solutions of the Einstein eqs. with the warped metric
(\ref{warped}) to be consistent the following local equality must take
place
\begin{equation}
 \left  ( A' e^{n A}\right) '=  {2\pi G_5\over 3} e^{n A}
  \left( T_\mu {}^\mu + (2n
-4)T_5 {}^5 \right)- {1-n\over 12} W^{n-2} R_g ~.
 \label{condition}
\end{equation}

If the internal space is not closed, this  condition may be still useful
since it shows that some combination of the energy-momentum tensor is a
total derivative when the Einstein equations are satisfied. With proper
care
of
the boundary conditions   it provides a consistency condition.

From now on we will concentrate on investigation of compact internal
spaces  without boundary (as in cases of a spatially periodic
metric \cite{RSI}).   In this case the integral of the LHS  vanishes.
This gives a constraint
\begin{equation}
\oint e^{n A} \left( T_\mu {}^\mu + (2n -4)T_5 {}^5 \right)=  {1-n\over
8\pi G_5} R_g \oint  e^{(n-2)A} ~,
 \label{constr}
\end{equation}
which can be used for all brane worlds with a spatially
 periodic metric. In case
of a flat brane, when $R_g=0$ the RHS of the eq. (\ref{constr})
vanishes.
The most interesting cases are $n=0,1,4$. For $n=0$
\begin{equation}
\oint  \left( T_\mu {}^\mu  -4T_5 {}^5 \right)=  {1\over 8\pi G_5} R_g
\oint  e^{-2A} ~.
 \label{constr0}
\end{equation}
This equation turns out to be particularly useful  for the study of  the
fine tuning required in the stabilization of the distance between the
branes. For $n=1$
\begin{equation}
\oint e^{ A} \left( T_\mu {}^\mu -2 T_5 {}^5 \right)= 0~.
 \label{constr1}
\end{equation}
The right-hand-side  of this constraint  vanishes in the special case
that $n=1$ even
for the non-flat brane! The constraint  with $n=1$ for flat branes with
$R_g=0$ has been  derived previously\footnote{In \cite{Kanti} a
different constraint is suggested: the dependence on energy-momentum tensor is as
in our eq. (\ref{constr1})  but the dependence on the metric is
different, which seems to contradict \cite{ellwanger} as well as our
constraints.} in \cite{ellwanger}. However it was suggested there that
this constraint is
necessary for the vanishing of the four-dimensional cosmological
constant.
As we see here, this particular constraint at $n=1$ does not
differentiate a flat from a non-flat brane. For $n=4$
\begin{equation}
\oint e^{4 A} \left( T_\mu {}^\mu + 4T_5 {}^5 \right)= - {3\over 8\pi
G_5}
R_g \oint  e^{2 A}= - {3\over 8\pi G_4} R_g ~.
 \label{constrNEW}
\end{equation}
This constraint was derived and analysed before in \cite{peter}. It was shown
 there that
it is satisfied in the RS model \cite{RSI}.

It will be useful the give an explicit form of the energy-momentum
tensor
for the class of theories we will be looking at. The Einstein-Hilbert
action is
\begin{eqnarray}
  S_{EG} =\int_{M}  d^5x\sqrt{-G} \left ( 2 M^3 R - {g_{IJ} \over 2}
   \partial_M \Phi^I \partial^M \Phi^J  - V(\Phi)-
  \sum_{\alpha} \lambda_\alpha(\Phi) \delta(y-y_\alpha)\right),
\label{EG}
\end{eqnarray}
where $4 M^3 =(8\pi G_5)^{-1}$. In the Randall-Sundrum scenario
\cite{RSI} the mass parameter $M$ is not much different from the 4d
Planck mass $M_p$. The interesting combinations of the
energy-momentum tensor are
\begin{equation}
  T_\mu^\mu = -4 \left({1\over 2} \Phi'\cdot \Phi' + V(\Phi) +
  \sum_{\alpha} \lambda_\alpha(\Phi) \delta(y-y_\alpha) \right),
\label{Tmumu}
\end{equation}
and
\begin{equation}
  T_5^5 =  {1\over 2} \Phi'\cdot \Phi' - V(\Phi)~.
\label{T55}
\end{equation}
Here the $\cdot$ indicates a contraction with the curved metric on the
scalar moduli
space.
From equations of motion we find that $2M^3 R = -{1\over 3} T_M^M$.

For this class of actions the basic local constraint takes the form
\begin{eqnarray}
&& -\left  ( A' e^{n A}\right)'=  {1-n\over 12} e^{(n-2)A} R_g
\nonumber\\
 && +  {1\over 12\cdot 4 M^3} e^{n
A}
  \left( (4-n) \Phi'\cdot \Phi'   + 2n V(\Phi) +4
   \sum_{\alpha} \lambda_\alpha(\Phi) \delta(y-y_\alpha)\right).
 \label{expl}
\end{eqnarray}
For compact internal space  without boundary the integral of the
left-hand side
vanishes and the constraint integrated over the period takes the form
\begin{eqnarray}
 &&
\oint   e^{n A}
  \left( (4-n) \Phi'\cdot \Phi'   + 2n V(\Phi) +4
   \sum_{\alpha} \lambda_\alpha(\Phi) \delta(y-y_\alpha)\right)
 \nonumber\\ &&
+4 M^3 \oint (1-n) e^{(n-2)A} R_g =0 ~.
 \label{expPeriodic}
\end{eqnarray}
There are three particular cases where this constraint may be especially
useful.

For $n = 1$ one has a condition which is to be satisfied independently
of the internal curvature $R_g$ of the brane (i.e. independently of the
effective four-dimensional cosmological constant):
\begin{eqnarray}
\oint  e^{ A}
  \left(  3  \Phi'\cdot \Phi'   + 2 V(\Phi) +4
   \sum_{\alpha} \lambda_\alpha(\Phi) \delta(y-y_\alpha)\right) =0 ~.
 \label{expPeriodic1}
\end{eqnarray}
For $n = 4$ this gives a constraint that does not include $\Phi'\cdot
\Phi'$:
\begin{eqnarray}
 \oint e^{4 A}
  \left(2 V(\Phi) +
   \sum_{\alpha} \lambda_\alpha(\Phi) \delta(y-y_\alpha)- 3 M^3 e^{2 A}
R_g \right) = 0~.
 \label{expPeriodic4}
\end{eqnarray}
For $n=0$ one finds another constraint, which will be particularly
useful for testing the consistency of various models:
\begin{equation}
 \oint  \left(
\Phi'\cdot \Phi'
 + \sum_{\alpha} \lambda_\alpha(\Phi) \delta(y-y_\alpha)+ M^3 e^{-2 A}
R_g \right)=0~.
 \label{con}
\end{equation}
It is important to realize that this condition does not
depend upon the choice of the  potential
$V(\phi)$, which  drops out from this combination. We may rewrite it as
follows:
\begin{equation}
 \sum_{\alpha} \lambda_\alpha(\Phi_\alpha)  + \oint
\Phi'\cdot \Phi'
= - M^3 R_g \oint e^{-2 A}   \ .
 \label{constraint}
\end{equation}
If the  geometry on the brane is of  de Sitter type, one has $R_g>0$  in
our notation. In this case one has
\begin{equation}
 \sum_{\alpha} \lambda_\alpha(\Phi_\alpha)  + \oint
\Phi'\cdot \Phi'  < 0 \ .
 \label{deSitter}
\end{equation}
If one wants to describe our universe, where $R_g$  vanishes with
an accuracy of about $10^{-120} M_P^2$, one can assume, with this
accuracy,
that $R_g =0$. In this case our consistency condition (\ref{deSitter})
becomes
\begin{equation}
 \sum_{\alpha} \lambda_\alpha(\Phi_\alpha)  + \oint
\Phi'\cdot \Phi'  =0 \ .
 \label{deSitter2}
\end{equation}
In what follows we will use the last two consistency conditions.

\section{Applications of consistency conditions}

\subsection{Randall-Sundrum two-brane scenario}

In the case of the original Randall-Sundrum scenario with two singular
branes and vanishing cosmological constant on the brane, $R_g=0$,  and
without scalar fields \cite{RSI} this constraint reads
\begin{equation}
 \lambda_h = -\lambda_v\, .
\label{theorem0}
\end{equation}
Here $\lambda_h$ and $\lambda_v$ correspond to the tension of the hidden
and visible branes respectively. The consistency condition is indeed
satisfied in \cite{RSI}, the tensions are constant and have  opposite
signs. This is a fine tuning which is in agreement with our consistency
constraint.

If one substitutes the metric with $A = -k|y|$ and $V=\Lambda$ into our constraint (\ref{expPeriodic1}), one finds the second RS condition
\begin{equation}
\Lambda  = k \lambda_v\, .
\label{theorem01}
\end{equation}
This condition represents the second fine tuning required in the RS scenario.

On the other hand, our constraint  (\ref{deSitter2}) shows that  a {\it
 smooth generalization of the flat ($R_g=0$) periodic 2-brane
Randall-Sundrum
solutions
\cite{RSI} or 3-brane KMPRS solutions \cite{KMPRS}  without
singular
sources does not exist}.
Indeed at $\sum_\alpha \lambda_\alpha
(\phi(y_\alpha))=0$ the constraint is reduced to
\begin{equation}
\oint   \Phi'\cdot \Phi' = 0\label{theorem1} ~.
\end{equation}
This is not possible since the LHS is positive and the RHS is strictly
zero. The constraint is satisfied only for constant scalars which
cannot build a smooth brane.

For $R_g > 0$ the corresponding condition becomes $\oint   \Phi'\cdot
\Phi'  < 0$, which is even more difficult to satisfy.

A similar no-go theorem has been established in \cite{KL} for the
non-singular supersymmetric generalizations of the one-brane
Randall-Sundrum scenario \cite{RSII}. As we see, for the two-brane
scenario the no-go theorem can be established independently of the issue
of supersymmetry. Thus in what follows we will assume that the branes
are singular.

\subsection{ Goldberger-Wise mechanism of brane stabilization}
  In presence of scalars and singular sources the constraint
(\ref{deSitter2}) shows
that {\it the total tension of the branes must be strictly negative and
equal to
the volume integral of the gradient energy of the scalars with the
opposite
sign}:
\begin{equation}
- \sum_\alpha \lambda_\alpha (\Phi(y_\alpha))=  \oint   \Phi'\cdot \Phi'
>0~.
\label{theorem2}
\end{equation}
We would like to check whether this consistency condition is satisfied
in existing models of singular branes on orbifolds stabilized in the
presence
of scalars.

In the original version of the two-brane  Randall-Sundrum
scenario, the distance
between
the branes could take any value. Goldberger and Wise proposed a simple
stabilization mechanism for the distance between the branes \cite{GW}.
They
added to the original Randall-Sundrum  model with brane tensions
$\lambda_h =
-\lambda_v$ a scalar field $\Phi$ with
bulk action
\begin{equation}
S_b={1\over 2}\int d^4 x\int_{-\pi}^\pi dy\, \sqrt{G}\,
\left(G^{AB}\partial_A \Phi \partial_B \Phi - m^2 \Phi^2\right),
\end{equation}
where $G_{AB}$  is the metric in the Randall-Sundrum scenario. They
also included interaction terms on the hidden and visible branes (at
$y=0$ and $y=\pi$ respectively) given by
\begin{equation}
S_h = -\int d^4 x \sqrt{-g_h}\, \gamma_h \left(\Phi^2 - v_h^2\right)^2,
\qquad
S_v = -\int d^4 x \sqrt{-g_v}\, \gamma_v \left(\Phi^2 - v_v^2\right)^2,
\end{equation}
where $g_h$ and $g_v$ are the determinants of the induced metric on the
hidden and visible branes respectively, and $\gamma_h$ and $\gamma_v$
are some {\it positive} coupling constants. An investigation of this
system
for large positive constants $\gamma_v$ and $\gamma_h$ has shown that
$\Phi(0) \approx v_h$, $\Phi(\pi) \approx v_v$, and the distance $r$
between the branes becomes stabilized at
\begin{equation}
\label{eq:min} r_c \approx \left(\frac{4}{\pi}\right) \frac{k}{m^2}
\ln\left[\frac{v_h}{v_v}\right].
\end{equation}
A certain advantage of this mechanism is that the brane stabilization occurs even if one changes a bit the bulk cosmological constant in the RS scenario. Thus at least one of the two fine tunings of the RS model can be relaxed.

However, a detailed investigation of the brane stabilization was performed in \cite{GW} without taking into account the back reaction of the scalar field on the metric. Our consistency conditions in the form (\ref{deSitter}), (\ref{deSitter2})   allow us immediately to check whether this approximation is consistent.

To use eq.  (\ref{deSitter2}) in any model of the brane world
with branes located at the fixed points of an ${S^1/ {\bf Z}_2} $
orbifold all we have to know is that the  {\it total negative tension of
the branes compensates exactly the positive integral of the gradient
terms of scalars}.
In the Goldberger-Wise model our constraint (\ref{deSitter2}) takes
the form
\begin{eqnarray}
\int _{-\pi}^{\pi}dy\; \Phi'\cdot \Phi' \, + \, \left[ \lambda_{h} +
\lambda_{v}
+\gamma_h \left(\Phi^2(0) - v_h^2\right)^2+\gamma_v \left(\Phi^2(\pi) -
v_v^2\right)^2\right] = 0 ~ . \label{GW}
\end{eqnarray}

In the original version of the RS  model  without stabilization
\cite{RSI}
one needs to impose the fine tuning condition $\lambda_{h} +
\lambda_{v}=0$. Thus our
constraint (\ref{deSitter2}) was satisfied. The Goldberger-Wise
mechanism (in the simplest approximation when the backreaction is neglected)
presumes that the condition $\lambda_{h} + \lambda_{v}=0$ still holds. But
then the constraint cannot be satisfied because  there are three new {\it
positive} contributions
which do not cancel: the term $\int _{-\pi}^{\pi}dy\; \Phi'\cdot
\Phi'$ and the terms $\gamma_h \left(\Phi^2(0) - v_h^2\right)^2$ and
$\gamma_v \left(\Phi^2(\pi) - v_v^2\right)^2$.

The term $\int _{-\pi}^{\pi}dy\; \Phi'\cdot \Phi'  $ is strictly
positive because the field $\Phi$ changes from (approximately) $v_h$ to
$v_v$, and the distance $r_c$ is nonzero only if $v_h \not = v_v$. Thus
$\Phi' \not = 0$, and $\int _{-\pi}^{\pi}dy\; \Phi'\cdot
\Phi'  > 0 $. One could expect that the terms $\gamma_h
\left(\Phi^2(0) - v_h^2\right)^2$ and $\gamma_v \left(\Phi^2(\pi) -
v_v^2\right)^2$ may vanish. However, the field $\Phi$ is not {\it
exactly}
equal to $v_h$ and $v_v$ on the branes. One can show that in the large
coupling constant approximation used in \cite{GW} one has $\gamma_h
\left(\Phi^2(0) - v_h^2\right)^2 = {m^4\over 64 k^2 \gamma_h}$, and
$\gamma_v \left(\Phi(\pi)^2 - v_v^2\right)^2= {m^4\over 64 k^2
\gamma_v}$. These terms vanish in the limit $\gamma_h, \gamma_v \to
\infty$, in agreement with \cite{GW}. However, for any finite $\gamma_h$
and $\gamma_v$ these terms are positive.
Thus all three terms discussed above are strictly positive, which
violates the consistency condition (\ref{GW}). 

Goldberger and Wise discussed the possibility to  alter the Randall-Sundrum condition $\lambda_{h} + \lambda_{v}=0$ (\ref{GW}) in order to find a self-consistent solution. Our constraint implies that to achieve consistency one should take $\lambda_{h} + \lambda_{v}<0$.  One may argue that a small change of brane tensions should not destabilize the system, but it should affect $R_g$. Then by making $\lambda_{h} + \lambda_{v}$ slightly negative one will still have a stable two-brane system. By gradually decreasing $\lambda_{h} + \lambda_{v}$ one may eventually achieve the desirable regime $R_g = 0$ for some value of $\lambda_{h} + \lambda_{v}$.

This would imply a more elaborate fine tuning of $\lambda_{h} + \lambda_{v}$ than in the RS model, where one simply required  $\lambda_{h} = - \lambda_{v}$. Here one would need to find a solution of a combined system of equations for the metric and for the scalar field, and only after that one would know whether a consistent solution with flat branes exists and what kind of fine tuning is actually required. 
To check the consistency of the solution one would need to verify not only the simplest constraint (\ref{constraint}), but also (\ref{expPeriodic1}), as we did for the RS model.

\newpage

\subsection{DeWolfe-Freedman-Gubser-Karch
 stabilization mechanism}
 
A consistent realization of the  Goldberger-Wise scenario for a certain class of potentials of the bulk scalars was proposed by
DeWolfe, Freedman, Gubser, and Karch  \cite{DFGK}.
In their model the first brane is
placed at $r=0$, the second brane at $r=r_0$, and the point $2r_0$ is
identified with the point $r=0$ so that $\int_{0}^{2r_0}dr \;
\Phi'\cdot \Phi' =2\int_{0}^{r_0}dr \; \Phi'\cdot \Phi'$. The
constraint (\ref{constraint}) takes the form
\begin{equation}
 \lambda_1 (\Phi_1) +
\lambda_2 (\Phi_2) = -2\int_{0}^{r_0}dr \; ( \Phi'\cdot \Phi'  + M^3
e^{-2 A} R_g ) ~. \label{theoremF}
\end{equation}
The solution for the flat branes with $R_g=0$ has the following property:
$\Phi' = {1\over 2} {\partial
W \over
\partial \Phi}$ in the volume between the first and the second brane.
The scalar dependent tensions of the brane are chosen as follows\footnote{The relation between the brane tension and the superpotential which allows the jump conditions to be satisfied  was first suggested in \cite{SB} in the context of $N=8$ gauged supergravity.}:
\begin{equation}
\lambda_1(\Phi_1) = W(\Phi_1) \ ,  \qquad \lambda_2(\Phi_2) =- W(\Phi_2)
\label{finetuning}\end{equation}
The
potential is presented via a `superpotential' $W$ as $V(\Phi)= {1\over
8}\left({\partial W \over
\partial \Phi}\right)^2 - {1\over 3} W(\Phi)^2$.  The choice of the
`superpotential' $W(\Phi) = {3\over L} - b\Phi^2$ is not given by any
known supersymmetric theory. However, this choice is useful for
providing an explicit example of the  flat brane solution with the
vanishing cosmological constant\footnote{The issue of the stabilization
of the distance between the branes is more subtle than in \cite{GW} as it 
requires an additional assumption on the relation between $\lambda_i''(\Phi_i)$ and
$W''(\Phi_i)$.}.
Performing the integral in eq.
(\ref{theoremF}) one finds that the constraint  requires that
\begin{equation}
b(\Phi^2_1 - \Phi_2^2) + \lambda_1 (\Phi_1) + \lambda_2 (\Phi_2)=0 ~.
\label{theoremF1}
\end{equation}
Here $br_0= \ln (\Phi_1/\Phi_2)$. The distance between the branes $r_0$
is positive and therefore for $b>0$, $\Phi_1> \Phi_2$ and for $b<0$,
$\Phi_2> \Phi_1$. This explains the  fine tuning condition $\lambda_1
(\Phi_1) + \lambda_2 (\Phi_2)= W(\Phi_1) - W(\Phi_2)$  required in
\cite{DFGK} for the vanishing of the cosmological constant. Indeed the
positive contribution from the volume integral $b(\Phi^2_1 - \Phi_2^2)>0
$ is precisely cancelled by the negative total tension of the branes
$\lambda_1 (\Phi_1) + \lambda_2 (\Phi_2)= b(\Phi^2_2 - \Phi_2^1)$.
We checked that other consistency conditions are also satisfied in this model.

Since the theory proposed in \cite{DFGK} is not supersymmetric, one should
take special care to ensure vanishing (or smallness) of the cosmological
constant with an account taken of radiative corrections.

\subsection{Supersymmetric branes in singular spaces}
As we see,  simple consistency checks can be very helpful and can bring
unexpected results when analyzing popular versions of the brane world
scenario. So far, we have applied these checks to the simplest
five-dimensional non-supersymmetric models. For a while it was not clear
whether a supersymmetric brane world theory is possible at all. Until
very
recently, a complete supersymmetry had not been formulated in singular
spaces, so all attempts have been  focused on the possibility of
obtaining
branes as non-singular domain wall solutions of supergravity
theories. However,
after a series of no-go theorems \cite{KL}, the attention shifted
towards the possibility of  formulating five-dimensional supergravity
in
singular spaces. The first example of such supersymmetric theory
involving a scalar and a vector fields was proposed in \cite{BKP}
\footnote{The domain wall solution  of  Horava-Witten theory
\cite{HoravaWitten}  reduced to 5 dimensions in \cite{LukOvrSteWal:1998}
satisfies the consistency condition (\ref{constraint}). The complete
supersymmetry  of the embedding theory is, however, not fully
understood.}. Consistency of this theory was achieved without  fine
tuning, but in a very nontrivial way. It was necessary to extend the
gauged five-dimensional supergravity to include the five-form field
strength $F= dA+\dots$
which on shell is  piece-wise constant. The gauge coupling constant
$g$  was promoted to a field $G(x)$. When the equations of motion from
the
bulk and the brane actions are solved for the potential $A$, the gauge
coupling field   becomes, on shell,  piece-wise constant, changing the
sign across the wall: $G(x)= g \epsilon (x)$.  Branes are sources of
five-form fluxes which are absent in the standard formulations of the
five-dimensional gauged supergravities.

The supersymmetric brane action depends on the induced metric
$g_{(4)}$, on the superpotential of the scalar fields $W$ and on  the
projection of the potential $A_{\mu \nu \rho\sigma} $  in the brane
directions:
  \begin{equation}
{\cal L}_{brane}=-2 g \,\left( \delta (x^5)-\delta
(x^5-\tilde x^5 )\right)
  \left( \sqrt g_{(4)} 3 W(\phi) +{1\over 4!} \varepsilon ^{\mu \nu \rho
\sigma }A_{\mu \nu \rho
  \sigma }\right) .
\label{SbraneA}
\end{equation}
 This bosonic action  is supersymmetric because the
variation of the two terms combines into a fermionic field which
vanishes on the brane as the result of ${\bf Z}_2$ symmetry. Thus the
form of the brane action and of the brane tensions are determined by
supersymmetry.

Here we will make an additional consistency check of this theory using
our methods. It was shown in \cite{BKP} that the bulk and brane
supersymmetry with the orbifold condition requires that the tensions of
the branes are related to the superpotential as follows:
$\lambda_1(\Phi_1) = 3 g W(\Phi_1)$, $\lambda_2(\Phi_2) =- 3 g
W(\Phi_2)$. This condition, up to a normalization, is exactly the
fine tuning condition (\ref{finetuning})  of Ref. \cite{DFGK}.  The main
difference is that in \cite{BKP} this condition appears automatically as a result of supersymmetry in singular spaces.

We can calculate the volume integral of the gradient energy using the
BPS relation between the derivative of the scalars and the derivative of
the superpotential with respect to  the scalars, which leads to
\begin{equation}
   \Phi'\cdot \Phi' =    3  G(x) W' ~.
\label{BPS}
\end{equation}
Thus the integral of the gradient energy of scalars is given precisely
by $3 W(\Phi_2) - 3 W(\Phi_1)$ which cancels the expression for the
tensions in the constraint equation (\ref{constraint}). Because of the
BPS
property of the solution which has 1/2 of the unbroken supersymmetry and
is derived in the framework of the supersymmetric bulk and brane action,
the brane cosmological constant in (\ref{constraint})  has to vanish.
Thus, not surprisingly, supersymmetry in the present context requires
a flat brane solution without any need for fine tuning.

\

In conclusion, we have derived some simple consistency conditions for
brane solutions in spatially periodic internal space for which the
integrals of total derivatives vanish. The simplest one is given in eq.
(\ref{constraint}). It relates the tensions on two branes to the volume
integral of the gradient energy of scalars and the cosmological constant
on the brane. 

\

The authors are grateful to  D. Freedman and G. Horowitz for valuable
discussions and encouragement and to  S. Gubser, W.~D.~Goldberger and
M.~B.~Wise for very useful correspondence and important comments. This work
was supported in part by NSF grant PHY-9870115.


\newpage


\begin{thebibliography}{30}

\bibitem{savas} N. Arkani-Hamed, S. Dimopoulos and G. Dvali,
 ``The hierarchy problem and new dimensions at a millimeter,'' Phys.
Lett. {\bf B429}, 263 (1998),   hep-ph/9807344;   I.~Antoniadis,
N.~Arkani-Hamed, S.~Dimopoulos and G.~Dvali,
``New dimensions at a millimeter to a Fermi and superstrings at a TeV,''
Phys.\ Lett.\  {\bf B436}, 257 (1998), hep-ph/9804398.


\bibitem{RSI} L.~Randall and R.~Sundrum,
 ``A large mass hierarchy from a small extra dimension,''
Phys.\ Rev.\ Lett.\  {\bf 83}, 3370 (1999)
[hep-ph/9905221].
%%CITATION = HEP-PH 9905221;%%

%\cite{Gibbons:1985}:
\bibitem{gary}
G.~W.~Gibbons,
 ``Aspects of Supergravity Theories,'' in {\it Supersymmetry,
Supergravity and Related
Topics}, eds. F. del Aguila, J. A. de Azcaragga and L. E. Ibanez,
World Scientific (1985).

\bibitem{ellwanger}
%\cite{Ellwanger:2000pq}:
U.~Ellwanger,
 ``Constraints on a brane-world from the vanishing of the cosmological
constant,''
Phys.\ Lett.\  {\bf B473}, 233 (2000)
[hep-th/9909103].
%%CITATION = HEP-TH 9909103;%%




\bibitem{peter}
S.~Forste, Z.~Lalak, S.~Lavignac and H.~P.~Nilles,
``The cosmological constant problem from a brane-world perspective,''
JHEP {\bf 0009}, 034 (2000)
[hep-th/0006139].


\bibitem{GW} W.~D.~Goldberger and M.~B.~Wise,
 ``Modulus stabilization with bulk fields,''
Phys.\ Rev.\ Lett.\  {\bf 83}, 4922 (1999)
[hep-ph/9907447].
%%CITATION = HEP-PH 9907447;%%

\bibitem{DFGK}
O.~DeWolfe, D.~Z.~Freedman, S.~S.~Gubser and A.~Karch,
 ``Modeling the fifth dimension with scalars and gravity,''
Phys.\ Rev.\  {\bf D62}, 046008 (2000)
[hep-th/9909134].
%%CITATION = HEP-TH 9909134;%%


%\cite{Bergshoeff:2000zn}:
\bibitem{BKP}
E.~Bergshoeff, R.~Kallosh and A.~Van Proeyen,
 ``Supersymmetry in singular spaces,''
JHEP {\bf 0010}, 033 (2000)
[hep-th/0007044].
%%CITATION = HEP-TH 0007044;%%


\bibitem{Kanti}
P.~Kanti, I.~I.~Kogan, K.~A.~Olive and M.~Pospelov,
``Single-brane cosmological solutions with a stable compact extra  dimension,''
Phys.\ Rev.\  {\bf D61}, 106004 (2000)
[hep-ph/9912266];  I.~I.~Kogan and G.~G.~Ross,
 ``Brane universe and multigravity: Modification of gravity at large
and  small distances,''
Phys.\ Lett.\  {\bf B485}, 255 (2000)
[hep-th/0003074].


\bibitem{KMPRS} I.~I.~Kogan, S.~Mouslopoulos, A.~Papazoglou, G.~G.~Ross
and J.~Santiago,
 ``A three three-brane universe: New phenomenology for the new
millennium?,''
Nucl.\ Phys.\  {\bf B584}, 313 (2000)
[hep-ph/9912552].

\bibitem{KL}
R.~Kallosh and A.~Linde,
``Supersymmetry and the brane world,''
JHEP {\bf 0002}, 005 (2000) [hep-th/0001071];
%%CITATION = HEP-TH 0001071;%%
%\href{\wwwspires?eprint=HEP-TH/0001071}{SPIRES}
K.~Behrndt and M.~Cvetic, ``Anti-de Sitter vacua of gauged
supergravities
with 8 supercharges,'' Phys.\ Rev.\  {\bf D61}, 101901 (2000)
[hep-th/0001159];
%%CITATION = HEP-TH 0001159;%%
%\cite{Gibbons:2000hg}:
G.~W.~Gibbons and N.~D.~Lambert,
``Domain walls and solitons in odd dimensions,''
Phys.\ Lett.\  {\bf B488}, 90 (2000)
[hep-th/0003197];
%%CITATION = HEP-TH 0003197;%%
%\href{\wwwspires?eprint=HEP-TH/0001159}{SPIRES}
J.~Maldacena and C.~Nunez,
``Supergravity description of field theories on curved manifolds and a no  go theorem,''
hep-th/0007018.
%%CITATION = HEP-TH 0007018;%%

%\cite{Randall:1999vf}:
\bibitem{RSII}
L.~Randall and R.~Sundrum,
``An alternative to compactification,''
Phys.\ Rev.\ Lett.\  {\bf 83}, 4690 (1999)
[hep-th/9906064].
%%CITATION = HEP-TH 9906064;%%


\bibitem{SB}
A.~Brandhuber and K.~Sfetsos,
%``Non-standard compactifications with mass gaps and Newton's law,''
JHEP {\bf 9910}, 013 (1999)
[hep-th/9908116].


\bibitem{HoravaWitten}
P.~Ho\u{r}ava and E.~Witten,
``Eleven-dimensional supergravity on a
manifold with boundary,''
Nucl.\ Phys.\  {\bf B475} (1996) 94
[hep-th/9603142].
%%CITATION = HEP-TH 9603142;%%
%\href{\wwwspires?eprint=HEP-TH/9603142}{SPIRES}



\bibitem{LukOvrSteWal:1998}
A.~Lukas, B.~A.~Ovrut, K.~S.~Stelle and D.~Waldram,
``Heterotic M-theory in five dimensions,''
Nucl.\ Phys.\  {\bf B552}, 246 (1999)
[hep-th/9806051].
%%CITATION = HEP-TH 9806051;%%


\end{thebibliography}
\end{document}